\documentclass[fleqn,10pt]{wlscirep}

\usepackage[utf8]{inputenc}
\usepackage[T1]{fontenc}

\usepackage{graphicx}
\usepackage{multirow}
\usepackage{color}
\usepackage{bm}
\usepackage{amsfonts,amssymb,amsmath}
\usepackage{graphicx,dcolumn,bm,color,braket,slashed}
\usepackage{times}
\usepackage{comment}
\usepackage{cleveref}
\usepackage{pdfpages}

\def\t#1{\textrm{#1}}

\title{
Symmetry-Protected Solitons and Bulk-Boundary Correspondence
in Generalized Jackiw-Rebbi Models}
\author[1+]{Chang-geun Oh}
\author[2+]{Sang-Hoon Han}
\author[1,2,3*]{Sangmo Cheon}
\affil[1]{Research Institute for Natural Sciences, Hanyang University, Seoul, 04763, Korea}
\affil[2]{Department of Physics, Hanyang University, Seoul, 04763, Korea}
\affil[3]{Institute for High Pressure, Hanyang University, Seoul, 04763, Korea}

\affil[+]{these authors contributed equally to this work}
\affil[*]{sangmocheon@hanyang.ac.kr}

\begin{abstract}
We investigate the roles of symmetry and bulk-boundary correspondence in characterizing topological edge states in generalized Jackiw-Rebbi (JR) models.
We show that time-reversal ($T$), charge-conjugation ($C$), parity ($P$), and discrete internal field rotation ($Z_n$) symmetries protect and characterize the various types of edge states such as chiral and nonchiral solitons via bulk-boundary correspondence in the presence of the multiple vacua.
As two representative models, we consider the JR model composed of a single fermion field having a complex mass and the generalized JR model with two massless but interacting fermion fields.
The JR model shows nonchiral solitons with the $Z_2$ rotation symmetry, whereas it shows chiral solitons with the broken $Z_2$ rotation symmetry.
In the generalized JR model, only nonchiral solitons can emerge with only $Z_2$ rotation symmetry, whereas both chiral and nonchiral solitons can exist with enhanced $Z_4$ rotation symmetry.
Moreover, we find that the nonchiral solitons have $C, P$ symmetries while the chiral solitons do not, which can be explained by the symmetry-invariant lines connecting degenerate vacua.
Finally, we find the symmetry correspondence between multiply-degenerate global vacua and solitons such that ${T}$, ${C}$, ${P}$ symmetries of a soliton inherit from global minima that are connected by the soliton, which provides a novel tool for the characterization of topological solitons.
\end{abstract}
\begin{document}

\flushbottom
\maketitle
\thispagestyle{empty}
\section*{Introduction}

The bulk-boundary correspondence and symmetry play pivotal roles in understanding topological materials.
The bulk-boundary correspondence is that those of the bulk govern the topological properties of the edge modes, and it has been confirmed in many topological materials\cite{hasan_colloquium_2010,qi_topological_2011}. 
Symmetry classifies and protects the topological properties of such materials. For instance, topological insulators and superconductors are classified by 10-fold Altland-Zirnbauer classes \cite{altland_nonstandard_1997,schnyder_classification_2008,ryu_topological_2010}, based on time-reversal (${T}$), charge-conjugation (${C}$), and chiral symmetries, and topological crystalline insulators are protected by space-group symmetries \cite{ando_topological_2015,hsieh_topological_2012,fang_new_2015,fu_topological_2011}. These studies strongly imply that analyzing the bulk and symmetry is important to understand the physical properties of topological edge states.
 
The Jackiw-Rebbi (JR) and Su-Schrieffer-Heeger (SSH) models are the famous 1D models presenting such topological behaviour \cite{jackiw1983,jackiw_solitons_1981,jackiw_solitons_1976,niemi_fermion_1986,heeger_solitons_1988,su_solitons_1979}.
The JR model describes a massless fermion field interacting with a nontrivial soliton background bose field in one spatial dimension \cite{jackiw1983,jackiw_solitons_1981,jackiw_solitons_1976}.
These models exhibit soliton's exotic topological properties such as zero-energy mode, a half fermion number, and spin-charge separation only when the nontrivial topology of bulk is protected by either ${C}$ or parity $(P)$ symmetry. 
This pioneering study is an example of the general principle that symmetry leads to richer topology and has stimulated many studies:
conducting polymer \cite{heeger_solitons_1988}, diatomic chain \cite{rice_elementary_1982}, magnetic system \cite{bluhm_dephasing_2011}, photonic crystal systems \cite{zhou_dynamically_2017, zhao_topological_2018,st-jean_lasing_2017}, artificially engineered SSH lattices \cite{huda_tuneable_2020, queralto_topological_2020}, ultracold atomic systems\cite{atala2013direct,nakajima2016topological,lu2016geometrical,leder2016real,meier2016observation} and so on.
Though there were several attempts to extend the SSH and JR models \cite{rice_elementary_1982,Vayrynen2011,li2014,shiozaki2015,zhao2016,zhang2017,velasco2017,xie2019,cheon_chiral_2015,han_topological_2020,oh_particle-antiparticle_2021, go_solitons_2013},
there have been only a few limited studies about the roles of symmetry and bulk-boundary correspondence that distinguish the nature of topological solitons beyond the SSH and JR models.

Chiral edge states are ubiquitous in nature 
and often protected by topology and symmetry \cite{chiu_classification_2016,hasan_colloquium_2010,qi_topological_2011}. 
Such chiral topological excitations are expected to show promise for information technology due to their robustness against external perturbations \cite{braun_topological_2012,parkin_magnetic_2008,romming_writing_2013}, and  had been treated as the hallmark of 2D and 3D topological insulators \cite{hasan_colloquium_2010,qi_topological_2011}.
Recently, topological solitons having chirality or chiral solitons were experimentally realized in a 1D electronic system with $Z_4$ topological symmetry \cite{cheon_chiral_2015,kim_topological_2012}.
Chiral switching using such solitons was demonstrated, which implies that chiral and nonchiral solitons are expected to be used as multi-digit information carriers \cite{kim_switching_2017}.
However, the universal conditions for the emergence of chiral and nonchiral solitons in 1D are not disclosed yet.

The purpose of this Letter is to discover the role of symmetry and bulk-boundary correspondence in characterizing various topological solitons in 1D systems using JR models.
In this work, we show that the cooperation of time-reversal, charge-conjugation, parity, and discrete field rotation ($Z_n$) symmetries protects and identifies the various types of soliton states via bulk-boundary correspondence.
We consider two representative models: The JR model, composed of a single fermion field with a complex mass, possesses phase rotation symmetry. The generalized JR model with two massless but interacting fermion fields has a discrete internal field rotation symmetry. 
The JR model shows charge-conjugate and parity-invariant solitons (equivalently, nonchiral solitons) in the presence of $Z_2$ field rotation symmetry, otherwise it shows chiral solitons with broken $C$ and $P$ symmetries.
The generalized JR model allows more abundant kinds of solitons---nonchiral (NC), right chiral (RC), and left chiral (LC) solitons [Fig.~\ref{fig:fig2}(d)]---depending on the controllable physical parameters.
In the generalized JR model, the $Z_2$ rotation symmetry protects NC solitons imposing the equivalence between an NC soliton and its anti-NC soliton.
The enhanced $Z_4$ rotation symmetry promotes the symmetry of global minima from $Z_2$ to $Z_4$, which endows the chirality to solitons.
Therefore, we reveal that the internal symmetry enriches the variety of topological solitons in 1D systems.
Further ${T}$, ${C}$, ${P}$ analysis discloses that a pair of chiral solitons forms a particle-antiparticle pair, while a nonchiral soliton is its antiparticle.
Finally, we discover the symmetry correspondence between multiply-degenerate global vacua and solitons such that ${T}$, ${C}$, ${P}$ symmetries of a soliton inherit from global minima that are connected by the soliton.

\section*{Results and Discussion}
\begin{figure}[t]
\centering
\includegraphics[width=130mm]{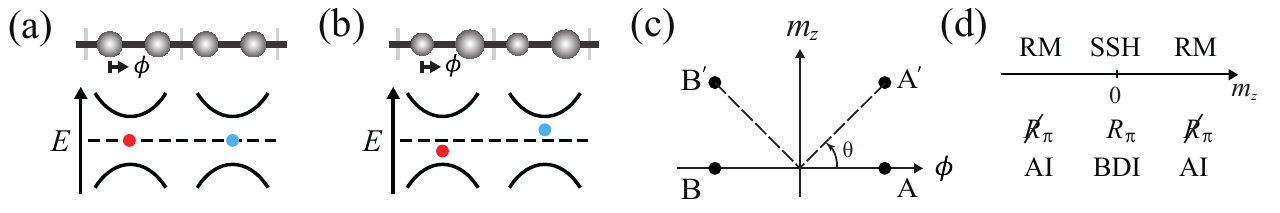}
\caption{
\label{fig:fig1}
Schematics of atom configurations and band structures for (a) SSH and (b) RM models.
$\phi$ is a dimerization displacement that corresponds to a bose field $\phi(x)$ in JR model.
Red and blue circles represent isolated soliton and antisoliton modes in the gap, respectively.
(c) The degenerate vacua for SSH and RM models in the order parameter space of $(\phi, m_z)$:
$A$ and $B$ ($A'$ and $B'$) for the SSH (RM) model. 
(d) Atland-Zirnbauer class \cite{altland_nonstandard_1997} and $\pi$ rotation symmetry with respect to the fermion mass $m_z$. 
$R_{\pi}$ ($\slashed{R}_{\pi})$ indicates that the $\pi$ rotation symmetry is preserved (broken).
}
\end{figure}

The low-energy effective theory of the SSH ($m_z = 0$) and the Rice-Mele (RM, $m_z \neq 0$) models \cite{su_solitons_1979, rice_elementary_1982} is described by the Lagrangian density of the JR model \cite{jackiw_solitons_1976} with a fermion mass $m_z$ [Fig.~\ref{fig:fig1}(a,b)]:
    \begin{eqnarray} \label{eq:JR_Lag_1}
        \mathcal{L}_{\text{JR}} =   \bar{\psi}[i\gamma^{0} \partial_0 + i \gamma^1 \partial_1 - \phi(x)-i \gamma^5 m_z]\psi,
    \end{eqnarray}
where $\psi(x)$ and $\phi(x)$ are two-component spinor and bose fields, and  $\gamma^0 = \sigma^y, \gamma^1 = -i\sigma^z, \gamma^5 =  \sigma^x$.
The mass terms can be represented as $ \phi(x) + i \gamma^5  m_z = m(x) e^{i \gamma^5 \theta(x)}$, where $m(x)=\sqrt{\phi^2(x) + m_z^2}$ and $\tan \theta (x) = \frac{m_z}{\phi(x)}$.
For a homogeneous bose field, $\phi(x)=\phi_0$, this model has gapped spectra of $E=\pm \sqrt{k_x^2+ \phi_0^2+m_z^2}$ and two degenerate vacua.
In the complex plane, the two degenerate vacua are located at $\theta$ and $\pi-\theta$ with $\tan \theta = \frac{m_z}{\phi_0}$,
which are denoted as $A$ and $B$ ($A'$ and $B'$) for the SSH (RM) model [Fig.~\ref{fig:fig1}(c)].
If the bose field $\phi(x)$ becomes a solitonic background field, $\phi(x)= \phi_s(x)$, that interpolates the two degenerate vacua, there appears an isolated soliton mode in the gap.
Depending on $m_z$, the emergent soliton modes are different.
The soliton modes in the SSH model are zero-energy modes, while the RM type soliton modes are not [Fig.~\ref{fig:fig1}(a,b)]. 

The JR model in Eq.~(\ref{eq:JR_Lag_1}) has the following global rotation symmetry:
    \begin{eqnarray} \label{rotation_RM_3}
        \theta \to \theta + \tilde \theta,~~~
        \psi(t,x) \to U \psi = i \exp \left(-i \gamma^5 \frac{ \tilde \theta }{2}\right)\psi(t,x).
    \end{eqnarray}
Because we are interested in the rotation symmetry that exchanges the two global minima, we set $\tilde \theta = \pi - 2\theta$.
The rotation symmetry in Eq.~(\ref{rotation_RM_3}) becomes $Z_2$ symmetry (equivalently, $\pi$ symmetry) only when $\theta=0$ (or $m_z=0$):
    \begin{eqnarray} \label{rotation_RM_4}
        \phi(x) \to e^{i\pi} \phi(x),~~~
        \psi(t,x) \to \gamma^5 \psi(t,x).
    \end{eqnarray}   
We also analyze ${T}$, ${C}$, and ${P}$ symmetries.
Because the JR model is spinless, ${T}$ symmetry exists.
On the other hand, the Lagrangian density in Eq.~(\ref{eq:JR_Lag_1}) has ${C}$ and ${P}$ symmetries only when $m_z=0$:
    \begin{eqnarray}
        \label{seq:JR_C_prove_2}
        && {C} \psi (t,x) {C}^{-1}= \psi^*(t,x), \\
		&& {P} \psi (t,x) {P}^{-1} = \gamma^0 \psi(t,-x),~~~~
		{P}\phi(x){P}^{-1} = \phi(-x).
	\end{eqnarray}
Therefore, $Z_2$ rotation (or $\pi$ rotation) symmetry protects the ${C}$ and ${P}$ symmetries, and vice versa.
Moreover, when the bose field is a soliton field, $ \phi(x) = \phi_s (x) $, the soliton system---composed of $\psi_s(x)$ and $\phi_s(x)$---is self-charge conjugate and parity-invariant\cite{jackiw1983,jackiw_solitons_1981,jackiw_solitons_1976}.
Thus, the JR model shows charge-conjugate and parity-invariant solitons (or, equivalently, nonchiral solitons) in the presence of the enhanced $Z_2$ rotation symmetry, otherwise it shows chiral solitons with broken $C$ and $P$ symmetries.

\begin{comment}
We also prove the particle-antiparticle duality between soliton and antisoliton using the charge-conjugation operation. 
The charge conjugated and parity transformed fields in a soliton system can be described by the fields in an anti-soliton system, and vice versa:
    \begin{eqnarray}
        \label{seq:JR_C_prove}
        && \psi^C_s =C \psi_s C^{-1}= -\gamma^5 \psi_{\bar{s}}^*, \\
		&& {P}\psi_{s}(t,x){P}^{-1} = \gamma^0 \gamma^5\psi_{\bar{s}}(t,-x),~~~~
		{P}\phi_s(x){P}^{-1} = \phi_s(-x) = \phi_{\bar{s}}(x).
	\end{eqnarray}
Such particle-antiparticle duality between soliton and anti-soliton systems appears in physical observables such as energy spectra [Fig.~\ref{fig:fig1}(b)].
\end{comment}
    
The solitons in the JR model can also be distinguished by topology.
With ${T}$, ${C}$, ${P}$ and $Z_2$ symmetries, 
the two minima in the JR model are distinguished by their quantized Berry phases; hence, the topological phase transition between the two minima occurs through the SSH type soliton.
With broken ${C}$, ${P}$, and $Z_2$ symmetries, in the RM model, the topology of the two minima can be continuously connected due to the fermion mass; thus, the soliton does not involve such phase transitions. 
Thus, an enhanced rotation symmetry can endow richer topological structures to solitons [Fig.~\ref{fig:fig1}(d)].

\begin{figure}[t]
\centering
\includegraphics[width=100mm]{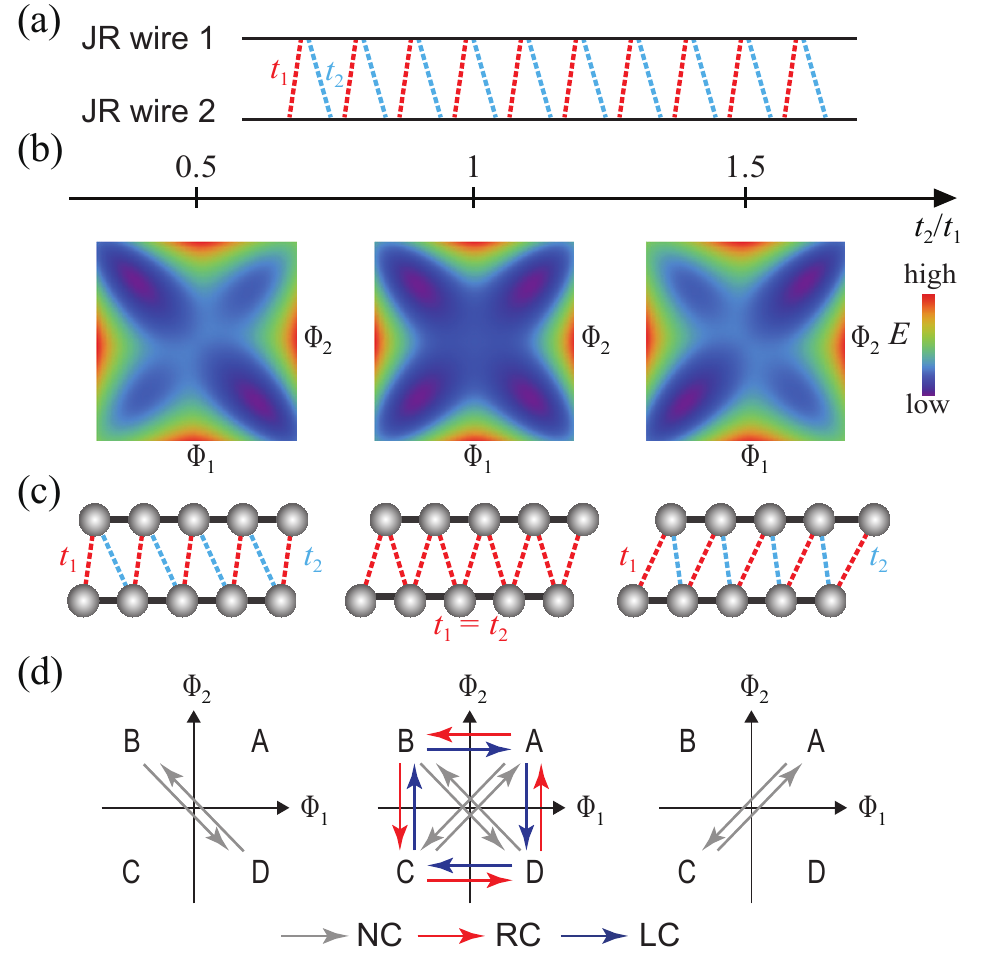} 
\caption{
\label{fig:fig2}
(a) Schematic for the generalized JR model composed of two JR wires and two interwire couplings ($t_1$ and $t_2$). In each $i$th JR wire, the $i$th fermion and bose fields, $\Psi_i(t,x)$ and $\Phi_i(t,x)$, reside. 
(b) Total energy profiles, (c) the corresponding lattice models, and (d) diagrams of possible solitons in the order parameter space ($\Phi_1,\Phi_2$) for three $t_2/t_1$ values. 
In (d), nonchiral (NC), right chiral (RC), and left chiral (LC) solitons are represented by grey, red and blue arrows, respectively. A, B, C, and D indicate global and local minima.
}
\end{figure}

Now, we consider the generalized JR model composed of two massless fermion fields and two interfield couplings ($t_1$, $t_2$).
The Lagrangian density is given by
	\begin{eqnarray}
		\mathcal{L} &=& \mathcal{L}_{1} + \mathcal{L}_{2} + \mathcal{L}_{\text{B}} + \mathcal{L}_{\text{int}} \label{tot_Lag},
		\\
		\mathcal{L}_j &=&   \bar{\Psi}_j[i\slashed{\partial} - \Phi_j ]\Psi_j,
		~~\mathcal{L}_{\text{B}} = - \frac{K}{2}(\Phi_1^2 + \Phi_2^2),
		\\
		\mathcal{L}_{\text{int}} &=& - \bar{\Psi}_1 [t_1 \gamma^0 - i t_2 \gamma^1]\Psi_2 - \bar{\Psi}_2 [t_1 \gamma^0 + i t_2 \gamma^1]\Psi_1, 
	\end{eqnarray}
where $\Psi_j$ and $\Phi_j$ are a two-component spinor field and a real scalar bose field localized in the $j$-th wire ($j = 1,2$), respectively. $\mathcal{L}_{\text{B}}$ corresponds to the potential of the bose fields, and $K$ is a positive constant.
Figure~\ref{fig:fig2} shows a schematic, the total energy profile, and solitons for several $t_2/t_1$ values.
When $t_1 \neq t_2$, there are two global and two local energy minima, and the atomic configuration in a global minimum has a staggered dimerization ordering.
As a realistic example, this case can describe the low-energy effective theory of the quasi-1D systems of polyacetylene chains \cite{fincher_structural_1982,baeriswyl_soliton_1983,baeriswyl_interchain_1988}.
At the critical point $t_1 = t_2$, the two local minima transform to global minima resulting in $Z_4$ symmetric four degenerate global minima. For example, this case can describe the low-energy effective theory of the coupled double Peierls chain model and explains the charge density wave system of indium atomic wires \cite{cheon_chiral_2015, yeom_instability_1999,bunk_structure_1999}.
The generalized JR model allows various kinds of solitons depending on $t_1$ and $t_2$.
When $t_1 \neq t_2$, only nonchiral solitons can exist, whereas chiral solitons can additionally exist at the critical point ($t_1=t_2$), as shown in Fig.~\ref{fig:fig2}(d).
The nature of chiral and nonchiral solitons will be explained shortly.

We find that the enhanced $Z_4$ rotation symmetry supports the emergence of chiral and nonchiral solitons.
Using Clifford matrices, the Lagrangian density in Eq. (\ref{tot_Lag}) can be rewritten as follows:
    \begin{eqnarray} \label{eq:Lag_03}
        \mathcal{L}= \bar{\Psi} [ \Gamma_2i \partial_0 +  \Gamma_3\partial_1 + t_1 \Gamma_{34} - i t_2 \Gamma_{25} -\Phi]\Psi + \mathcal{L}_\text{B}, ~~
    \end{eqnarray}
where $\Psi = (\Psi_1 , \Psi_2 )$, $\Phi = \left(\begin{array}{c c} \Phi_1 & 0  \\ 0 & \Phi_2  \end{array}\right)$,  $\bar{\Psi} = \Psi^\dagger \Gamma_2$ and
the form of Clifford matrices $\Gamma_i$ are given in Sec.~S2.1. in Supplemental Information.
Depending on interfield couplings, this Lagrangian density can have 
$Z_2$ rotation symmetry ($n=2$; $\pi$ rotation symmetry) or $Z_4$ rotation symmetry ($n=1$; $\pi/2$ rotation symmetry):
    \begin{eqnarray}
        \Psi(t,x) &\to& \left(\begin{array}{c c} 0 & -i \sigma_x  \\ 1 & 0 \end{array}\right)^n\Psi(t,x), \label{rotation_GJR}\\
        \Phi(x) &\to& e^{i\Gamma_{14} \theta}\Phi(x),
    \end{eqnarray}
where $ e^{i\Gamma_{14}\theta} = \left(\begin{array}{c c} \cos\theta & \sin \theta  \\ -\sin \theta & \cos \theta  \end{array}\right)$ and $\theta =\frac{n\pi}{2}$.
Similar to the JR model, the $Z_2$ rotation exchanges two global minima that are located oppositely with respect to the origin in order parameter space [Fig.~\ref{fig:fig2}(b)].
Hence, the $Z_2$ rotation symmetry implies that an NC soliton and its anti-NC soliton systems are identical, which indicates that the Lagrangian for an NC soliton system satisfies ${C}$ and ${P}$ symmetries as discussed later.
While the action is invariant under the $Z_2$ rotation regardless of $t_1$ and $t_2$, the $Z_4$ rotation symmetry is realized only at the critical point $t_1 = t_2$ [Sec.~S2.6. in Supplementary Information]. 
The $Z_4$ rotation rotates the energy profile counterclockwise by $\pi/2$ and hence the $Z_4$ symmetry guarantees four degenerate global minima, 
which indicates that the $Z_4$ rotation symmetry is necessary for the simultaneous emergence of the chiral and nonchiral solitons.
In addition, the $Z_4$ rotation symmetry supports the equivalence relations among the same type of  solitons because it changes a soliton to the soliton rotated counterclockwise by the same angle in order parameter space [Fig.~\ref{fig:fig2}(d)].

\begin{figure}[t]
\centering
\includegraphics[width=100mm]{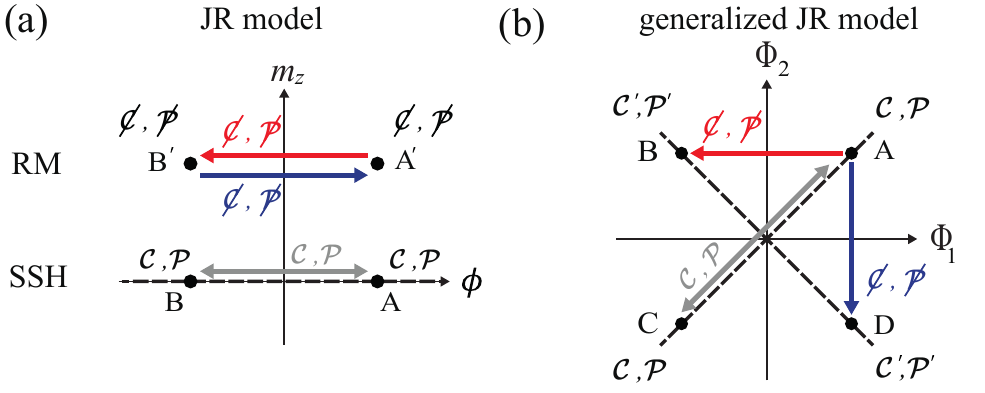} 
\caption{
\label{fig:fig3}
Diagrams for ${C}$ and ${P}$ symmetries of global minima and solitons for the (a) JR and (b) generalized JR models. ${C}$ and ${P}$ indicate the symmetries, while $\slashed{{C}}$ and $\slashed{{P}}$ do broken symmetries. The dashed lines represent ${C}$ and ${P}$ symmetry-invariant lines.
In (a), the $\phi$ axis and the symmetry-invariant line are overlapped.
The gray arrow represents solitons in the SSH model, and red and blue arrows do in the RM model.
$A$ and $B$ ($A'$ and $B'$) are two degenerate vacua for the SSH (RM) model. 
In (b), the prime in superscript represents that the corresponding symmetry transformation has a different form comparing with the unprimed one.
The red, blue, and gray arrows represent right-chiral, left-chiral, and non-chiral solitons, respectively.
}
\end{figure}

We analyze ${T}$, ${C}$, ${P}$ symmetries in the generalized JR model [see details in Sec.~S2.5 of Supplementary Information].
${T}$ symmetry is always present as the JR model is spinless.
On the other hand, ${C}$ and ${P}$ symmetries exist only along with the symmetry-invariant lines (AC and BD lines), and their forms are different depending on the lines [Fig.~\ref{fig:fig3}(b)].
The Lagrangian density in Eq.~(\ref{eq:Lag_03}) has the following symmetries:
For the AC line,
    \begin{eqnarray}
		{C}\Psi(t,x){C}^{-1} &=&-\Gamma_{14}\Psi^*(t,x),\\
		{P}\Psi(t,x){P}^{-1} &=& \Gamma_{34}\Psi(t,-x),
    \end{eqnarray}
and for the BD line,
    \begin{eqnarray}
		{C}\Psi(t,x){C}^{-1} &=&-\Gamma_{5}\Psi^*(t,x),\\
		{P}\Psi(t,x){P}^{-1} &=& \Gamma_{24}\Psi(t,-x). 
    \end{eqnarray}
In the viewpoint of energetics, only one symmetry-invariant line is realized when $t_1 \neq t_2$, whereas both symmetry-invariant lines are realized at the critical point $t_1 = t_2$. 
This indicates that the $Z_4$ rotation symmetry gives the coexistence of the two symmetry-invariant lines.

Now we unveil the symmetries of soliton systems by explicitly analyzing the Lagrangian density for soliton systems [see details in Sec.~S2.5 of Supplementary Information].
An NC soliton system has ${T},{C},{P}$ symmetries because the bose fields are localized in the symmetry-invariant line (AC or BD line).
This indicates that an AC soliton is its antiparticle having no chirality.
On the other hand, in RC and LC soliton systems, ${C}$ and ${P}$ symmetries are broken
because the bose fields deviate from the symmetry-invariant line, which naturally endows the chirality to solitons.
However, the charge conjugated and parity transformed fields of an RC soliton system can be described by the fields of an LC soliton system, and vice versa:
\begin{eqnarray}
		{C}\Psi_{\text{RC}(i \rightarrow j)}(t,x){C}^{-1} &=& -\Gamma_{14}U\Psi^*_{\text{LC}(j \rightarrow i)}(t,x),\\
  		{P}\Psi_{\text{RC}(i \rightarrow j)}(t,x){P}^{-1}
		&=& \Gamma_{34}U^*\Psi_{\text{LC}(j \rightarrow i)}(t,-x),
    \end{eqnarray}
where $i$ and $j$ in the subscript indicate two global minima that the corresponding soliton field connects, and the unitary matrices $U$ are given by
    \begin{eqnarray*}
        U=\left(\begin{array}{c c} 0 & 1  \\ -i\sigma_x & 0  \end{array}\right),~~ &\text{when}&~~ (i,j) = (A,B) ~\text{or}~ (C,D),\\
        U=\left(\begin{array}{c c} 0 & i\sigma_x  \\ 1 & 0  \end{array}\right),~~ &\text{when}& ~~(i,j) = (B,C) ~\text{or}~ (D,A).
    \end{eqnarray*}
This indicates that the RC and LC solitons are charge-conjugate and parity partner, i.e., they satisfy particle-antiparticle duality.
Similarly, in the JR model, particle-antiparticle duality between soliton and antisoliton can be proved [Sec.~S1.4. in Supplementary Information]. 
    
Similar to the bulk-boundary correspondence in topological insulators, we discover symmetry correspondences between multiple global minima and various types of solitons.
${T}$, ${C}$, ${P}$ symmetries of a soliton are determined by the symmetries of two global minima connected by the soliton. 
Thus, the symmetry properties of solitons can be obtained without definitive proof for the soliton system.
Figure~\ref{fig:fig3} shows the diagrams of ${T}$, ${C}$, ${P}$ symmetries for the JR and generalized JR models. 
In the SSH soliton system, two global minima are located in a symmetry-invariant line, and hence both two global minima and solitons connecting them have ${T}, {C}, {P}$ symmetries. 
In the RM soliton system, two global minima do not have ${C}$ and ${P}$ symmetries, hence the solitons do not have ${C}$ and ${P}$ symmetries. 
In the generalized JR model with the $Z_4$ rotation symmetry, each global minima have ${T},{C}$ and ${P}$ symmetries. 
However, the forms of ${C}$ and ${P}$ transformations are different according to the symmetry-invariant AC or BD lines, as discussed before.
When a soliton interpolates two global minima in the AC or BD line, the soliton has ${T},{C},{P}$ symmetries, which results in a self-charge-conjugated and parity-invariant NC soliton.
When a soliton connects two global minima out of the same symmetry-invariant line, the soliton does not have ${C}$ and ${P}$ symmetries, which endows chirality to the RC and LC solitons as shown in Fig.~\ref{fig:fig3}(b).

\begin{figure}[t]
\centering
\includegraphics[width=100mm]{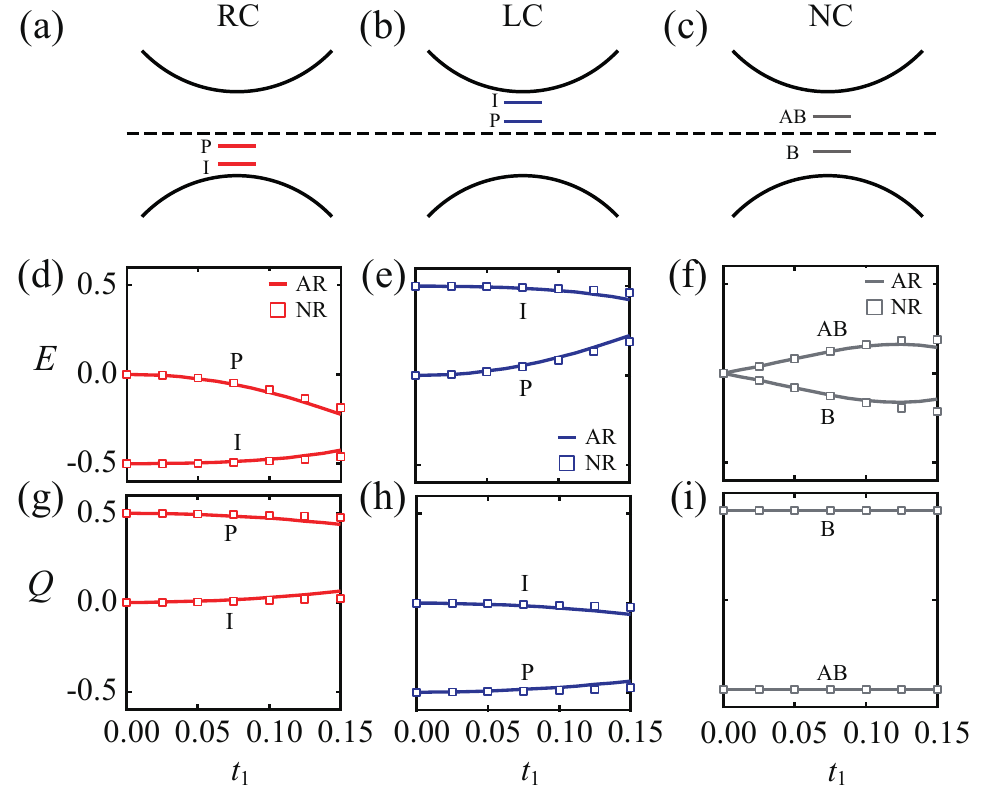}
\caption{
\label{fig:fig4}
Schematics for the soliton modes of (a) RC, (b) LC, and (c) NC soliton systems.
Energy spectra for soliton modes for (d) RC, (e) LC, and (f) NC systems with respect to $t_1$. The spectra are normalized by the energy gap for each $t_1$.
The fermion numbers for soliton modes with respect to $t_1$ for (g) RC, (h) LC, and (i) NC systems. P, I, AB, and B represent the primary, induced, antibonding, and bonding modes, respectively. Analytical results (AR; lines) and numerical results (NR; squares) are calculated at the critical point $t_1=t_2$. 
}
\end{figure}

Such ${C}$ duality between chiral soliton systems and ${C}$ symmetry in a non-chiral soliton system appear in physical observables such as energy spectra and fermion numbers of isolated soliton modes.
In the generalized JR model, two isolated soliton modes for each soliton system appear, and they are the results of fractionalization or hybridization of zero modes [Fig.~\ref{fig:fig4}].
For RC and LC soliton systems, one isolated energy mode is the primary mode, and the other is the induced mode [Fig.~\ref{fig:fig4}(a,b)].
When a soliton field exists in one fermion field resulting in the primary mode, an effective soliton field is induced in the other fermion field due to the interfield couplings, which leads to the emergence of the induced mode. Thus, a zero mode is fractionalized into primary and induced modes.
For an NC soliton system, the two isolated energy modes are distinguishable as bonding and antibonding modes regardless of $t_1$ and $t_2$ when $t_1 t_2 \neq 0$ [Fig.~\ref{fig:fig4}(c)]. They are the result of hybridization between two zero modes. 

We calculate the energy spectra and fermion numbers of such isolated soliton modes using the effective one-particle Hamiltonian and Goldstone-Wilczek method \cite{goldstone_fractional_1981}. See detailed calculations in Sec.~S2. of Supplemental Information.
Figure~\ref{fig:fig4}(d-f) shows the energy spectra with respect to the strength of the interwire coupling at the critical point ($t_1=t_2$).
The bonding and antibonding modes in an NC soliton system are symmetrically located with respect to $E = 0$ because of ${C}$ symmetry [Fig.~\ref{fig:fig4}(f)].
Because the nature of an NC soliton is protected by the $Z_2$ rotation symmetry and symmetry-invariant lines, the symmetric spectra of an NC soliton are maintained even if $t_1 \neq t_2$ [see Fig.~S1 in Supplementary Information].
However, primary and induced modes in each RC and LC soliton system are not symmetrically located because ${C}$ symmetry is broken.
Instead, the isolated energy modes in the RC soliton system are located symmetrically with the isolated energy modes in the LC soliton system [Fig.~\ref{fig:fig4}(d,e)].
This indicates that the RC and LC solitons form a charge-conjugate pair and satisfy $C$ duality.
Figure~\ref{fig:fig4}(g-i) shows the fermion number of each isolated mode with respect to the strength of the interfield coupling. 
We now consider the Fermi-level at $E=0$, i.e., the bonding mode in an NC soliton system and soliton modes in an RC soliton system are filled, while the other modes are empty. 
Similar to the energy spectra of RC and LC solitons, the fermion numbers of RC and LC solitons are also opposite [Fig.~\ref{fig:fig4}(g,h)].
The fermion number for the bonding and antibonding modes in an NC soliton system are also oppositely positioned, as shown in Fig.~\ref{fig:fig4}(i). This behavior is maintained even for $t_1 \neq t_2$ case [see Sec.~S2.3. in Supplementary Information].
Similarly, for the JR model, the spectra of isolated soliton and antisoliton modes show the $C$ duality and ${C}$ symmetry [Fig.~\ref{fig:fig1}(a,b)].
Therefore, chiral soliton shows ${C}$ duality, while nonchiral solitons do ${C}$ symmetry.

\section*{Conclusion}
In summary, we have studied the roles of symmetry and bulk-boundary correspondence in the JR and generalized JR models.
We showed that the cooperation between ${T}$, ${C}$, ${P}$, and internal discrete field rotation  symmetries characterizes the physical properties of topological solitons in 1D systems.
Furthermore, we found bulk-boundary correspondence between the multiply degenerate global minima and solitons such that the symmetry of a soliton inherits the symmetries of two global minima that the soliton interpolates.
Our work can be generally applied to 1D systems such as conducting polymers\cite{heeger1988solitons}, charge density wave systems\cite{gruner1988dynamics}, engineered atomic chain systems\cite{huda_tuneable_2020}, photonic crystal systems\cite{ozawa2019topological}, ultracold atomic systems\cite{cooper2019topological}, and higher dimensional systems.

\bibliography{main-arxiv}

\section*{Acknowledgements}
We thank T.-H.K. for valuable discussions.
C.-g.O., S.-H.H., and S.C. were supported by the National  Research  Foundation  (NRF)  of Korea  through  Basic  Science  Research  Programs (NRF-2018R1C1B6007607 and NRF-2021R1H1A1013517), the research fund of Hanyang University (HY-2017), and the POSCO Science Fellowship of POSCO TJ Park Foundation.

\section*{Author contributions statement}
S.C. conceived the project.
C.-G.O and S.-H.H performed theoretical and numerical calculations under S.C.
All authors discussed the results and co-wrote the manuscript.

\section*{Competing Interests} 
The authors declare no competing financial interests.

\section*{Additional information}

\noindent
\textbf{Correspondence} and requests for materials should be addressed to S.C.

\includepdf[pages=-]{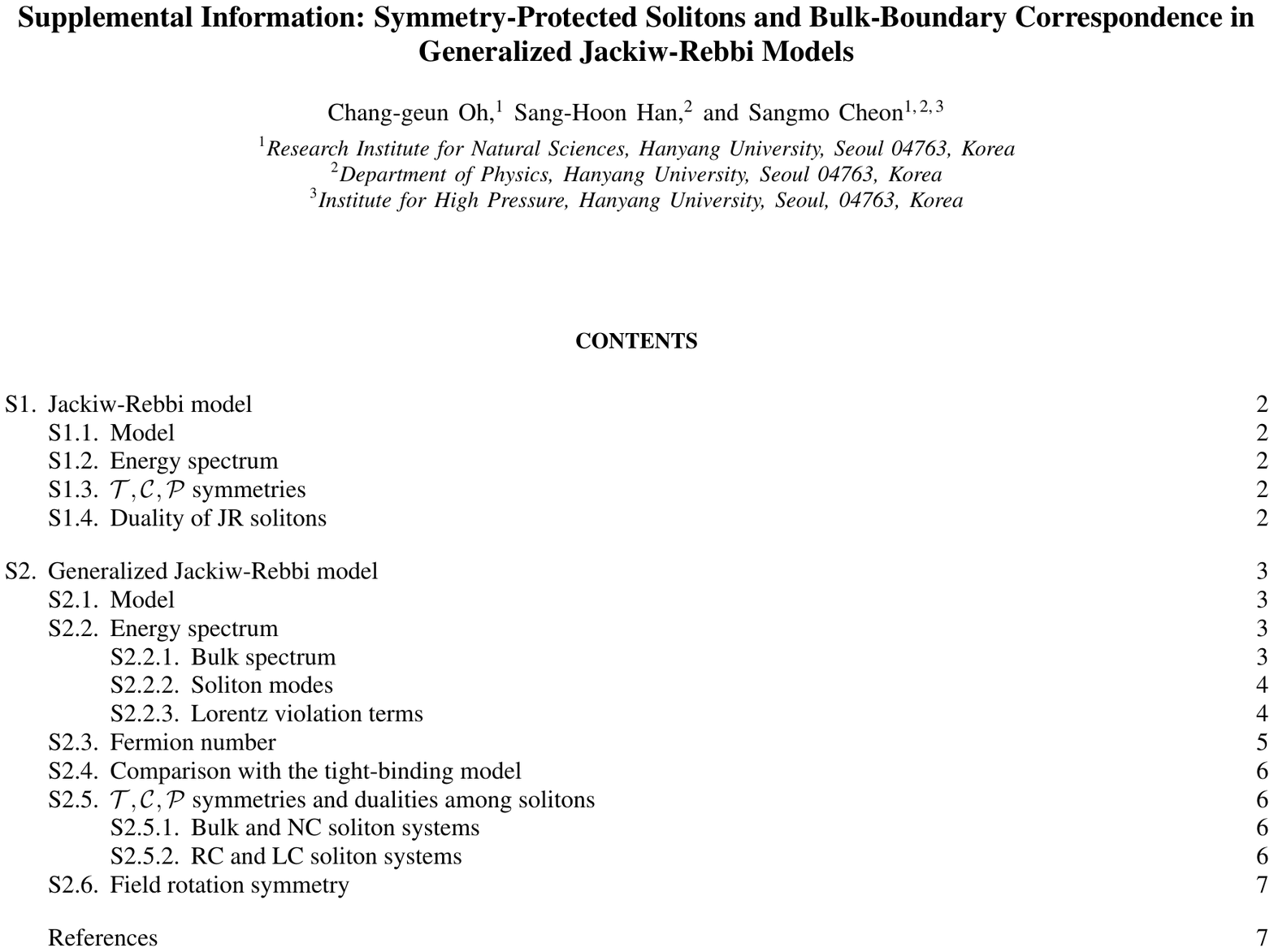}

\end{document}